\documentclass[prl,twocolumn,showpacs,floatfix,aps]{revtex4}
\usepackage{graphicx}

\begin{document}
\preprint{cond-mat/0000000}

\title{Phonon structure in I-V characteristic of MgB$_{2}$ point-contacts}

\author{I. K. Yanson$^{1,2}$\footnote{Corresponding author, e-mail:
yanson@ilt.kharkov.ua}, V. V. Fisun$^{1}$, N. L. Bobrov$^{1}$, and Yu. G. Naidyuk$%
^{1}$}

\affiliation{$^{1}$B.Verkin Institute for Low Temperature Physics
and Engineering, National Academy  of Sciences of Ukraine, 47
Lenin Ave., 61103,  Kharkiv, Ukraine}
\affiliation{$^{2}$Forschungzentrum Karlsruhe GmbH, Technik und
Umwelt, Postfach 3640, D-76021, Karlsruhe, Germany}
\author{W. N. Kang, Eun-Mi Choi, Hyun-Jung Kim and Sung-Ik Lee}

\affiliation{National Creative Research Initiative Center for
Superconductivity, Department of Physics, Pohang University of
Science and Technology, Pohang 790-784, South Korea}

\date{\today}

\begin{abstract}
The search of the  phonon structure at the above-gap energies was
carried out for $d^{2}V/dI^{2}(V)$ spectra of MgB$_{2}$ point
contacts with a normal metal. The two-band model is assumed not
only for the gap structure in $dV/dI(V)$-characteristics, but also
for phonons in $d^{2}V/dI^{2}(V)$ point-contact spectra, with up
to the maximum lattice vibration energy. Since the current is
carried mostly by charges of 3D-band, whereas the strong
electron-phonon interaction occurs in 2D-band, we observe the
phonon peculiarities due to ''proximity'' effect in {\it k}-space,
which depends both on the contact orientation with respect to the
crystal axes and the variation of interband coupling through the
elastic scattering.

\end{abstract}

\maketitle

{\it Introduction}.

The superconducting properties of the newly discovered {\it s-p} compound MgB$%
_{2} $ \cite{Akimitsu} have been attracting recently much
attention of the superconductive community. The relatively high
$T_{c}\simeq 39$ K, the absence of magnetic order and strongly
correlated charge carriers, on the one hand, and the strong links
between grains for large enough critical currents and magnetic
fields, on the other hand, make it very promising for complete
understanding of the superconducting mechanism and its successful
applications at moderate temperatures \cite{Buzea}.

Unfortunately, the simplicity of this compound does not go too far
for immediate understanding of all details of its superconducting
mechanism. From the theoretical point of view, this compound
appears to be a rare example of several (at least two)
disconnected bands of the Fermi surface (FS) with quite different
dynamical properties \cite{Shulga,Mazin,Liu}. The theory predicts
that one of those is two-dimensional (2D), with strong
electron-phonon interaction (EPI), while the other is 3D with weak
EPI and even larger density of states at the Fermi energy \cite
{Mazin,Liu,Yildirim}. The interband scattering between these two
parts is weak \cite {Liu}, and they behave like an ''intrinsic
proximity effect'' \ \cite{Schmidt}. The superconducting and
normal parts are not separated spatially, like in the McMillan
model \cite{McMillan,Wolf} of SN sandwiches, but are weakly
connected in {\it k}-space through the interband scattering. In
connection with this behavior, one has to recall earlier works of
two-band superconductivity \cite{Suhl,Sung,Galaiko}. But there is
an important difference between the predictions of those old
papers and what is observed in Ref.\cite{Schmidt} and some other
point-contact and tunneling works (see, for example,
Refs.\cite{Laube,Iavarone}). Since the 2D and 3D parts of Fermi
surface are very anisotropic, the tunneling and point-contact
current depends strongly on the orientation with respect to the
crystal axes, and for \textit{c}-orientation the contribution of
the 3D band prevails \cite{Brinkman}. Superconducting energy gap
values, corresponding to these two bands, differ very much (up to
the factor of 3 \cite{Brinkman,Choi}), and turn to zero at the
same critical temperature $\sim 38-40$ K. Moreover, if the
separate energy gaps converge to a single averaged gap, the
critical temperature remains almost the same. Possible explanation
of this behavior \cite{Brinkman}  is that the charges of the 3D
band play much greater role in the PC current, due to the higher
Fermi velocity and larger plasma frequency. By analogy, the
situation looks as if one measures superconducting energy gap only
from the "normal' side of the McMillan sandwiches in coordinate
space. $T_{c}$ does not decrease noticeably with increasing
elastic scattering, since the latter is not strong enough.

Luckily, in MgB$_{2}$ no other bosonic excitation, except phonons,
is anticipated \cite{Liu,Choi,Bohnen}. Thus, the only mechanism of
the formation of Cooper pairs
is EPI, although when a new superconductor with relatively high T$%
_{c}$ is discovered, other mechanisms are always proposed.

In spite of great variety of spectroscopic measurements, which
convincingly show the existence of two different superconducting
energy gaps belonging to the above mentioned two parts of FS (see,
for example, Refs.\cite{Laube,Naidyuk}, and more recent papers
\cite{Schmidt,Iavarone}), there are only a few reports of direct
experimental proof for phonon structure in the point-contact
characteristic \cite{Bobrov} or tunneling \cite{D'yachenko}.

In this paper we present the first attempt to identify the
structure of MgB$_{2}$ point-contact (PC) spectra  at energies up
to the maximum phonon frequency. We show that the observed
peculiarities are unequivocally caused by phonons, and their
appearance correlates with the observed superconducting energy gap
for a given point contact. In our opinion, this is due to the
strong anisotropy of multi-band electron and phonon spectra, and
to the influence of the elastic scattering, which determines both
the intraband and interband EPI coupling in this compound. The
random deviation of the contact axis  from the predominantly
c-direction plays a crucial role.

{\it Specific of experiment}

The samples are  {\it c}-axis oriented films of MgB$_{2}$, whose
characteristics are described in Refs.\cite{Naidyuk,Kang}. The
contacts are made by direct touch of MgB$_2$ with a sharpened
normal metal electrode (Cu or Ag).
The axis of the contact nominally coincides with the {\it c}%
-axis of the film, and thus, determines the main direction of the
current. The orientation selectivity in the point-contact
spectroscopy is low (lower, than in plane tunnel junction, where
it goes exponentially), and is within a solid angle of roughly a
few tens of degrees. Moreover, since not every touch produces the
desired appearance of the energy-gap I-V characteristic, we moved
the electrodes relative to each other in order to penetrate
through the accidentally damaged surface layer. This operation
changes the contribution of the perpendicular {\it ab}-direction
and produces additional defects in the contact region and, thus,
shortens further the small electron mean free path. In a few
experiments we tried to make a contact with a broken side-face of
the film, in order to increase the occurrence of {\it
ab}-direction for the current.

Thus, our ''input'' characteristic is the shape of superconducting
energy-gap structure, which is described elsewhere
\cite{Naidyuk,Bobrov}. The main goal is to investigate the first
and second derivatives of the I-V characteristic at large biases,
providing the contact survives the long-term measurements.

The attractive feature of PC spectroscopy is that the measured
second derivative for the normal state turns out to be directly
proportional to EPI function $\alpha^2(\omega)\,F(\omega)$
\cite{KOS,YansonSC}

\begin{equation}
\label{pcs} \frac{{\rm d}^2V}{{\rm d}I^2}\propto \frac{{\rm d}(\ln
R)}{{\rm d}V}= \frac{8\,ed}{3\,\hbar v_{\rm
F}}\alpha^2(\omega)\,F(\omega)
\end{equation}
where $\alpha$, roughly speaking, describes the strength of the
electron interaction with one or another phonon branch for
electron transport through the contact with diameter $d$,
$F(\omega)$ stands for the phonon density of states. In PC spectra
the EPI spectral function is modified by the transport factor,
which increases strongly the backscattering processes.

However, in the superconducting state the above-gap  singularity
at phonon frequencies can be caused by several mechanisms
\cite{YansonSC,Beloborod'ko}. The features, which are strongly
shifted along the voltage scale upon increasing the temperature
and field, should be disregarded, as due to the nonequilibrium
superconducting phenomena \cite{YansonSC,Beloborod'ko}.
Only those, whose positions do not depend on $%
T,H,$ except the small shifts of the order of superconducting
energy gaps (2$\div $7 meV), are considered.

The temperature is varied from 1.5 K up to $T\geq T_{c}$ with the
magnetic field up to 4.5 T applied along the {\it ab}-plane. At
low temperatures this field is not enough to destroy
superconductivity. In order to observe the inelastic point-contact
spectrum in the normal state, we raise both the temperature and
the field.

The magnetic field does not influence the inelastic PC spectrum
(1), because of the short mean free path. But it can strongly
suppress superconductivity in the weakly superconducting 3D band,
and destroy the exchange of the Cooper pairs from the strongly
superconducting 2D-band. Since the current flows mostly from the
3D-band, the phonon structure appears indirectly, its intensity
being strongly influenced by the field. Only small admixture of
the direct 2D-band contributes to EPI, due to random inclusion of
the {\it ab-}direction in the current.

{\it Results and Discussion}.

In Fig.1a the $dV/dI(V)$ and $V_{2}\left( eV\right)\propto
d^{2}V/dI^{2}\left( eV\right)$ characteristics are shown in the
superconducting state at zero field. In the
energy-gap region the two gaps are clearly seen with $\Delta _{1}=2.4$ and $%
\Delta _{2}=5.7$ meV. Further on, we simply denote by $\Delta $
the position of $dV/dI$ minimum. For many junctions we determine
the superconducting energy gap by the BTK-fitting procedure
\cite{Naidyuk}. For PC with not too large $\Gamma$-parameter,
introduced by Dynes, this value does not differ too much from the
true energy gap. The first derivative at above-gap energies is
slightly asymmetric and nearly parabolic, which leads to an
approximately linear background in $d^{2}V/dI^{2}(V)$. This
background is subtracted from the raw $V_2(V)$-dependence, and the
result is shown by curve 2 (panel a). The latter curve displays an
antisymmetric structure, whose amplitude is about a fraction of
percent in differential resistance. In the panel (b), this
structure is compared with the phonon density of states (PDOS)
\cite{Renker}. Good correspondence is seen between the positions
of PDOS peaks and maxima in $d^{2}V/dI^{2}(V)$.
\begin{figure}
\includegraphics[width=8cm,angle=0]{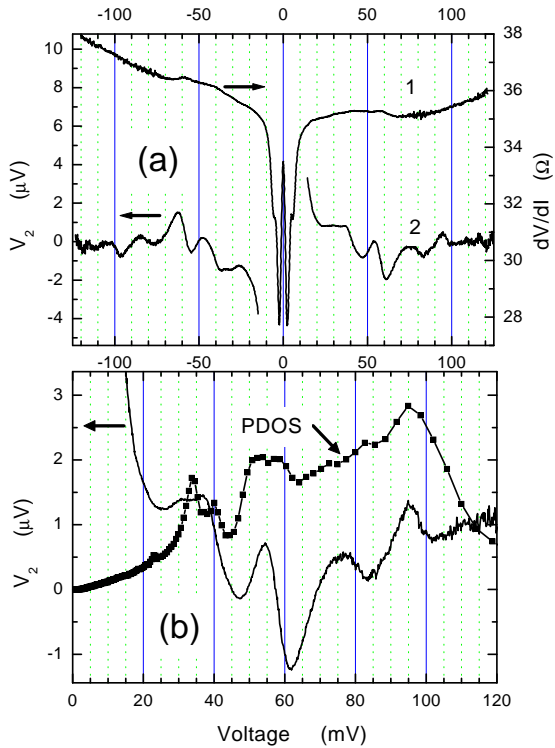}
\caption[]{a) {\bf Curve 1}: Differential resistance $R=dV/dI(V)$
for MgB$_{2}$-Ag point contact. The structure at bias <10 mV shows
two superconducting gaps corresponding to two different parts of
the Fermi surface. The junction is made with the broken side-face
of the film. {\bf Curve 2}: The point-contact spectrum $V_{2}
\propto d^{2}V/dI^{2}$ of the same junction ($V_{2}$ is the second
harmonic voltage with modulation
voltage $V_{1}=2.52$ mV).  $d(\ln R)/dV=2\sqrt{2}%
V_{2}/V_{1}^{2}$, where $R=dV/dI(V)$. The linear background is
subtracted from the raw data.

$T=4.2$ K, $H=0$, $R_{0}=35$ $\Omega$ .

b) The point-contact spectrum averaged between plus and minus
$V$-polarity (the same, as in panel a), compared to the
experimental phonon density of states of MgB$_{2}$ taken from
Ref.\cite{Renker}. The ordinate axis for PDOS is given in
arbitrary units.} \label{fig1}
\end{figure}

The reproducibility of this PC spectra for different contacts can
be seen in Fig.2a,b, where the raw data for three different
junctions are shown. Here, the contact resistance varies from 43
to 111 $\Omega $ with gap minima at $\Delta =$ 2.1, 2.6, and 4.86
meV, respectively. All the second derivative spectra (b) correlate
with PDOS. The slight variation is probably due to the anisotropy
and different scattering rates in the contacts. Compared with the
theoretical EPI function (see, for example, Ref.
\cite{Yildirim,Bohnen}), the peak at eV=60$\div $70 meV of
$E_{2g}$ boron mode is not too much higher in intensity, in accord
with observation of Ref.\cite{Bobrov}.

According to our classifications of the energy gap structure
\cite{Naidyuk}, it is due to the random orientation of the contact
axis and scattering of charge carriers between two band of Fermi
surface, having gaps $\Delta _{1}$ and $\Delta _{2}$ mixed. With
an increase of interband scattering the magnitude of $\Delta_1$
($\Delta_2$) moves to the higher (lower) value, respectively.  For
dirty contacts, where the admixture of the 2D-band is essential,
only one maximum remains with a broad distribution around
$\simeq3.5$ meV \cite{Naidyuk}, while $T_{c}$ remains almost the
same. In this case $l$ is smaller than $d$ (where $l$ and $d$ are
the electron mean free path and the size of the contact,
respectively), and the inelastic backscattering contribution to
the phonon structure, proportional to $l/d$ \cite{KulYan},  is
small. Therefore, in the superconducting state, the observed
phonon structure presents mainly the elastic contribution to the
excess current \cite {YansonSC}. The elastic term is proportional
to the energy dependent part of the excess current
$I_{exc}\left(eV\right)$, similar to the phonon structure in the
quasiparticle DOS for tunneling spectroscopy
\cite{YansonSC,Beloborod'ko}:

\begin{equation}
\frac{dI_{exc}}{dV}(eV)= \frac{1}{R_{0}}\left|\frac{\Delta
_{in}\left(
eV\right) }{eV+\sqrt{\left( eV\right)^{2} -\Delta _{in}^{2}\left( eV\right) }%
}\right|^{2} ,  \label{elastic}
\end{equation}
where $\Delta _{in}\left( eV\right) $ is the gap parameter in the
3D band, induced by 2D-band EPI. As seen in Fig.2, if the $\Delta$
value increases (panel a, curves 1-3), then the intensity
of the phonon structure in the units of $d(\ln R)/dV=2\sqrt{2}%
V_{2}/V_{1}^{2}$ increases too (panel b). That is because the
modulation voltage $V_1$ decreases, whereas the amplitudes in
$V_2(eV)$-units remain approximately the same. Note, that for
curve 1 in Fig.2a one has to take the lower gap $\Delta_1$,
because the current is mainly determined by 3D-band, as mentioned
above.
\begin{figure}
\includegraphics[width=8.5cm,angle=0]{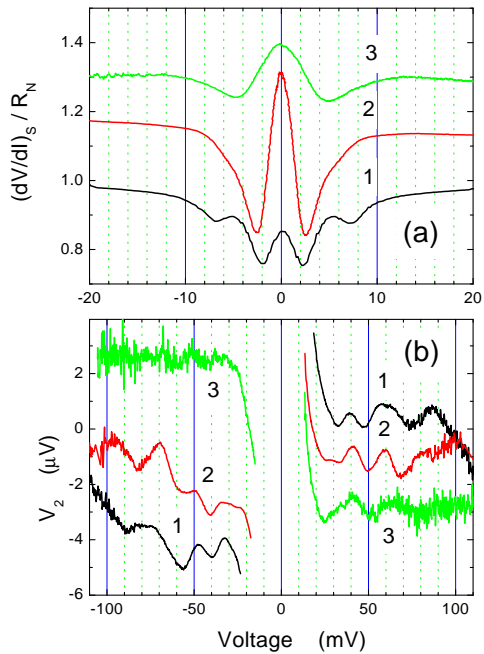}
\caption[]{(a) Differential resistance normalized to the normal
state (taken as a resistance at $V\simeq 30$ mV) for 3 different
contacts with $R_{0}$=45, 43, and 111 $\Omega $ (curves 1-3,
respectively). Main gap $dV/dI-$minima are at about 2.1, 2.6, and
4.86 mV, correspondingly. The larger gap for curve 1 equals 7.0
meV. Curves 2,3 are shifted vertically for clarity.

(b) Raw data of the second harmonic signal $V_2$ with modulation voltages $V_{1}=$%
3.31, 2.78, and 2.5 mV, respectively. The numbers of curves are
the same as in (a). $T=4.2$ K, $H=0.$ Curve 1 are shifted
vertically for clarity. Their center of symmetry has (0,0)
coordinates. } \label{fig2}
\end{figure}

In Fig.3, the normal state spectrum is displayed together with the
differential resistances in the superconducting state, showing the
energy gap structure. In spite of a 10 times increase of the
temperature smearing ($\simeq 20\,$\ meV at 41 K instead of 2 meV
at 4.2 K), the residual phonon structure is still visible in
Fig.3b. The smeared phonon features are superimposed on the rising
linear background. For curve 2, we see an increase in scattering
at $\simeq 35$ meV, where the acoustic phonon peak occurs, and the
saturation at $\simeq $100 meV, where the phonon spectrum ends. In
the normal state, only those nonlinearities remain, which are due
to the inelastic processes
 \cite{KOS}.  We stress that judging from the
superconducting gap value ($\simeq 3.69$ meV for curve 2 in panel
a) the essential contribution is expected from the $\Delta
_{2}(E)$ of 2D-band for normal-state spectrum No.2. This spectrum
should contrast with curve 6 in Fig.4 (see below), where the
direct contribution from the 2D-band is small, due to lower value
of $\Delta$. Thus, the backscattering processes from the 3D-band
are mostly essential here. The same is true for curve 6 in Fig.4,
where the phonon features are not seen at high energies. The shape
of spectrum 1 in Fig.3b presents the behavior intermediate between
these two extremes, although its energy gap is approximately equal
to curve 6 in Fig.4. Beside the well seen phonon spectral features
in the range $30 \div100$ meV, it possesses a low energy bump at
about 20 meV.

Note, that the modulation voltages for spectra 1,2 in Figs.3b and
4 are the same. That means that they have the same second
derivative scales, and the intensity of the second derivative is
greater for curve 2 in Fig.3b, than those of curve 1 in Fig.3b and
curve 6 in Fig.4.

\begin{figure}
\includegraphics[width=8.5cm,angle=0]{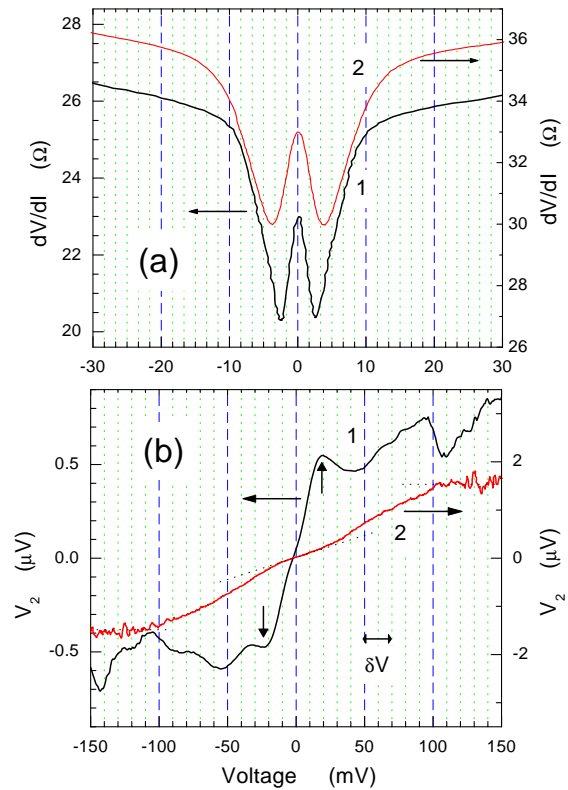}
\caption[]{(a) Differential resistances for MgB$_2$-Cu point contacts, whose
normal state spectra are presented in panel b. $\Delta^{(1)}=2.55$
meV, $\Delta^{(2)}=3.69$ meV. T=4.2 K, H=0. (b) Second harmonic
signals in the normal state. $V_{1}=2.2$ mV, $T=41$ K. Note the
different Y-scales for curves 1 and 2, which means that the
generation of phonons in the latter case is more intensive. For
spectrum No.1 the vertical arrows mark the low frequency bumps at
about 20 meV. The dotted tangent straight lines point to the
phonon features on curve No.2. The horizontal bar with double
arrows stands for thermal smearing, determining the spectral
resolution $\delta V$.} \label{fig3}
\end{figure}

The increasing second derivative of the I-V characteristic in the
normal state is due to excess generation of nonequilibrium phonons
at the contact. The PC spectra with phonon features in the normal
state at T$\geq$ T$_c$, like those shown in Fig.3b, usually
exhibit the nonequilibrium phenomena in the superconducting state.
These phenomena are mostly due to destruction of superconductivity
in the vicinity of the contact, and lead to depression of the
excess current \cite{YansonSC}. It has already been noted, that
the positions of these nonlinearities are strongly influenced by
the field and temperature, and should therefore be disregarded.

As to the spectacular electron-phonon spectrum, presented in our
preliminary report (see Fig.3,4 in Ref.\cite{Bobrov}), it is
presumably a rare example of a direct contribution from the
inelastic spectrum of the 2D-band. The magnetic field of 4 T
suppresses almost completely the induced phonon features from the
3D-band (with exception of low energy peak). Thus, the proposed
tentative consideration would bring our previous observations in
line with to what is consistently described here.

The phonon spectra of a contact with a small value of energy gaps
are characterized by the presence of low frequency phonon peaks.
The small peak at energy of about 25 meV (Fig.4, curves 1,2) is
visible, where a tiny knee exists on the PDOS \cite{Renker} (see
PDOS in Fig.1b). In the normal state (at $T\geq T_{c}$), these low
frequency peaks transform into the S-shape structure in
$d^{2}V/dI^{2}(V)$-spectra (curve 6), corresponding to the wide
minimum of $dV/dI(V)$ near zero bias. This low frequency structure
is hardly due to the remnants of superconducting quasigap at
$T>T_{c}$, since it is absent in junction No.2, whose
characteristic is shown in Fig.3. Rather, it could be thought of
as strong interaction in the 3D-band with very low-frequency
excitations, whose origin is not clear yet. On increasing the
temperature, the low frequency peak broadens further like common
spectral features do in the normal state.

Since most of the current through the point contact is carried by
the 3D band with weak EPI, the phonon singularity in the
superconducting state is induced mainly by interband scattering
from the 2D band with strong EPI. The degree of this induction
depends on the interband coupling. As was noted above, in the
clean limit the interband coupling in MgB$_{2}$ is low \cite
{Liu}. That was the reason for applying the
McMillan-proximity-effect model in order to fit the gap structure
in $dI/dV(V)$ characteristic in Ref. \cite{Schmidt}. This model
leads to the small-intensity phonon structure. The elastic
scattering, which is certainly high in thin film and may be even
stronger by point contact fabrication process, enhances the
interband, as well as the intraband coupling. Following Arnold
\cite{Arnold}, and assuming the extension of the proximity effect
theory in coordinate space to the two-band model (i.e. "proximity"
effect in $k$-space), we use the corresponding expression for the
interband coupling, induced by elastic scattering in SN
sandwiches. In our case, by analogy, the elastic scattering leads
to the expression

\begin{equation}
\Delta _{1}(E)=\frac{Z_{1}^{p}\Delta _{1}^{p}+i(\hbar /2\tau
)\Delta _{2}(E)N_{2}(E)}{Z_{1}^{p}(E)+i(\hbar /2\tau )N_{2}(E)},
\end{equation}
which, like the McMillan expression for $\hbar /2\tau =\Gamma
_{N}$ (not shown)$,$ derives to $\Delta _{1}(E)\simeq \Delta
_{2}(E)$, in other words, to the large induced phonon structure in
the 3D-band. Here $Z_{1}^{p}$ is
 the renormalization function in the 3D-band, $N_{2}(E)$ is the
 electron DOS in the 2D-band, $%
\tau $ is the elastic scattering time, and $\Gamma _{N}$ is the
interband scattering rate in the McMillan model. Thus, as the
interband scattering increases, the intensity of phonon
singularity moves to the higher energy, characteristic of the
2D-band. This ''internal proximity effect'' explains also the
small variation in the position (of the order of superconducting
energy gap) of phonon peaks with various elastic scattering rates
produced by contact fabrication.

\begin{figure}
\includegraphics[width=8cm,angle=0]{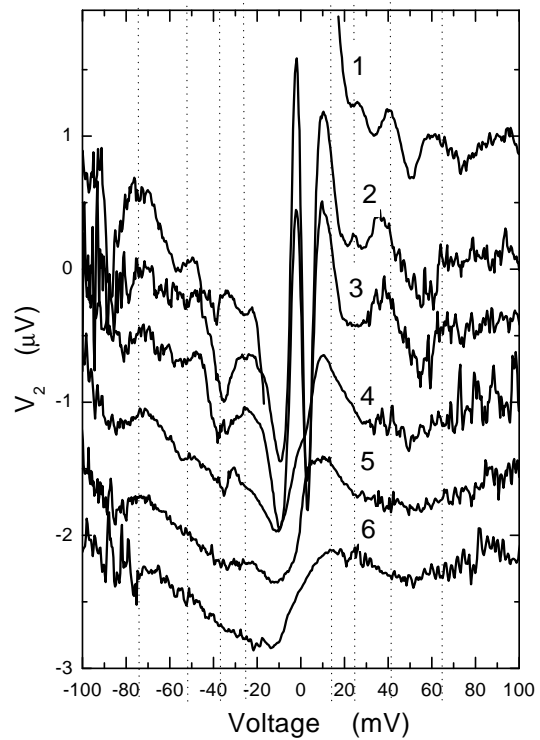}
\caption[]{Second harmonic dependences on field and temperature. The
temperature and magnetic field are equal to: 4.2 K, 0 T; 4.2 K, 4
T; 10 K, 4 T; 20 K, 4 T; 30 K, 4 T; 40 K, 0 T, for curves 1-6,
respectively. The modulation voltage $V_1$ is 2.2 mV.
$\Delta$=2.7 meV. The vertical dotted lines mark the phonon
singularity on the voltage axis. Curves are shifted vertically for
clarity, their center of symmetry corresponding to (0,0)
coordinates.} \label{fig4}
\end{figure}

The influence of magnetic field on the point-contact spectra is
shown in Fig.4 (compare curves 1 and 2). It is evident that the
field can smear out the intensity of high energy peaks, which are
induced in the 3D-band by EPI of the 2D-band. The disappearance of
phonon peaks at rising field and temperature proves that they do
not belong to the inelastic backscattering processes, which should
have the same intensity both in the superconducting and in normal
states. It seems more plausible that the high energy phonon peaks
are due to the elastic contribution in the excess current
(\ref{elastic}) induced by EPI from the 2D band, as was already
stated above.

Unfortunately, we cannot estimate the strength of EPI by inelastic
PC spectroscopy \cite{KOS}, since we are not able to destroy the
superconductivity completely by magnetic field at low enough
temperature. The latter is necessary to get good resolution. For
thin films this experiment will hardly promise encouraging
results, since the elastic mean free path is strongly reduced, and
the intensity of inelastic backscattering processes is therefore
diminished by the factor $l/d$. Note, that for the superconducting
self-energy effects the short elastic mean free path does not
suppress essentially the phonon structure \cite{YansonSC}. On the
contrary, for the two-band model elastic scattering even enhances
the phonon structure due to the ''intrinsic proximity effect''
described above.

{\it Conclusions}

We summarize our observation as follows.  The reproducible phonon
peaks are seen in the superconducting state on $d^{2}V/dI^{2}(V)$,
and disappear in the normal state, which proves that they are due
to the energy dependence of superconducting order parameter. The
superconductivity is presumably due to the 2D-band EPI, which
induces the $\Delta (E)$ structure in the 3D band. At a small
value of the superconducting gap the phonon structure is weak, and
its intensity begins to increase with growing $\Delta$. The
robustness against the magnetic field also grows notably with
increasing $\Delta$.

In the normal state the intensity of the inelastic spectrum also
correlates with the value of superconducting gap. The tendency is
observed that the high energy phonon peaks become more prominent
for a larger gap. For a smaller gap, the low-frequency bump
appears in the spectra.  It might be due to EPI with some unknown
low-frequency excitation \cite{Lampakis}, but the further work is
needed to be sure that it is not caused by phonon peaks of the
normal-metal counter electrode (Cu, Ag).

The final goal of PC spectroscopy is to use sufficiently clean
material at low temperatures in high enough magnetic field, in
order to suppress superconductivity and estimate quantitatively
the strength of EPI. The PC experiment, including single crystal
technology development, should go in parallel with theoretical
calculation, taking into account the specific transport
form-factor \cite{KOS}.

{\it Acknowledgements}

The work in Ukraine was supported by the State Foundation of
Fundamental Research, Grant F7/528-2001.

IKY is grateful to Forschungzentrum Karlsruhe for hospitality, and
thanks Dr. Renker for MgB$_2$ phonon spectrum from
Ref.\cite{Renker}.

The work at Postech
was supported by the Ministry of Science and Technology of Korea
through the Creative Research Initiative Program.


\begin{thebibliography}{}

\bibitem{Akimitsu}  J. Nagamatsu {\it et al., }Nature (London) {\bf 410%
}, 63 (2001)

\bibitem{Buzea}  D.M. Buzea and T. Yamashita, Supercond. Sci. Technol.
{\bf 14}, R115 (2001)

\bibitem{Shulga}  S.V. Shulga {\it et al., }cond-mat/0103154

\bibitem{Mazin}  J. Kortus {\it et al., }\ Phys. Rev. Lett {\bf 86},
4656 (2001)

\bibitem{Liu}  A.Y. Liu, I.I. Mazin, and J. Kortus, Phys. Rev. Lett. {\bf %
87, }087005 (2001)

\bibitem{Yildirim}  T. Yildirim {\it et al., } Phys.
Rev. Lett, {\bf 87}, 037001 (2001)

\bibitem{Schmidt}  H. Schmidt {\it et al., }\ Phys. Rev. Lett.
{\bf 88}, 127002 (2002)

\bibitem{McMillan}  W.L. McMillan, Phys. Rev. {\bf 175}, 537 (1968)

\bibitem{Wolf}  E.L. Wolf, {\it Principles of Electron Tunneling
Spectroscopy, }Oxford, 1985

\bibitem{Suhl}  H. Suhl, B.T. Matthias, and L.R. Walker, Phys. Rev. Lett.
{\bf 3}, 552 (1959)

\bibitem{Sung} C.C. Sung and V.K. Wong, J. Phys. Chem. Solids {\bf
28}, 1933 (1967)

\bibitem{Galaiko} V.P. Galaiko, Fiz. Nizk. Temp. {\bf 13}, 1102
(1987) [Sov. J. Low Temp. Phys. {\bf 13}, No.10 (1987)]

\bibitem{Laube} F. Laube, G. Goll, J. Hagel, H. v.
L$\ddot{o}$hneysen, D. Ernst, T. Wolf, Europhys. Lett. {\bf 56},
296 (2001)

\bibitem{Iavarone}  M. Iavarone {\it et al.}, cond-mat/0203329

\bibitem{Brinkman} A. Brinkman {\it et al.}, Phys. Rev. \textbf{B65}, 180517 (2002)

\bibitem{Choi}  H.J. Choi {\it et al., }cond-mat/0111182 and
cond-mat/0111183

\bibitem{Bohnen}  K.-P. Bohnen, R. Heid, and B. Renker, Phys. Rev. Lett.
{\bf 86}, 5771 (2001)

\bibitem{Naidyuk}  Yu.G. Naidyuk, I.K. Yanson {\it et al.}, JETP Letter
{\bf 75}, 238-241 (2002)

\bibitem{Bobrov}  N.L. Bobrov {\it et al., }cond-mat/0110006, to
be published in book {\it "New Trends in Superconductivity" },
eds. J.F. Annett and S. Kruchinin, Kluwer Academic Publishers,
Dodrecht, 2002

\bibitem{D'yachenko}  A.I. D'yachenko {\it et al.}, cond-mat/0201200

\bibitem{Kang} W. N. Kang, Hyeong-Jin Kim, Eun-Mi
Choi, C. U. Jung, Sung-Ik Lee, Science {\bf 292}, 1521 (2001);
W.N. Kang, H.-J. Kim, E.-M. Choi {\it et al.}, Phys. Rev. Lett.
{\bf 87}, 087002 (2001)

\bibitem{KOS}  I.O. Kulik, A.N. Omelyanchouk, and R.I. Shekhter, Sov. J. Low
Temp. Phys. {\bf 3, }840 (1977)

\bibitem{KulYan} I.O. Kulik and I.K. Yanson, Sov. J. Low Temp.
Phys. {\bf 4}, 596 (1978)

\bibitem{YansonSC}  I.K. Yanson, cond-mat/0008116; in book I.O. Kulik and
R.Ellialtioglu (eds.){\it , Quantum Mesoscopic Phenomena and
Mesoscopic Devices in Microelectronic,} p. 61-77, (2000), Kluwer
Academic Publishers.

\bibitem{Beloborod'ko}  A.N. Omel'yanchuk, S.I. Beloborod'ko and
I.O. Kulik, Sov. J. Low Temp. Phys. {\bf 14}, 630 (1988)

\bibitem{Renker}  B. Renker {\it et al.,} Phys. Rev. Lett {\bf 88},
067001 (2002)

\bibitem{Arnold}  G.B. Arnold, Phys. Rev. B{\bf 23}, 1171 (1981)

\bibitem{Lampakis} D. Lampakis, cond-mat/0105447

\end{thebibliography}
\end{document}